
%
%
%
%
\input harvmac.tex

\def\frac#1#2{{\textstyle{#1\over #2}}}

\def\b#1{\kern-0.25pt\vbox{\hrule height 0.2pt\hbox{\vrule
width 0.2pt \kern2pt\vbox{\kern2pt \hbox{#1}\kern2pt}\kern2pt\vrule
width 0.2pt}\hrule height 0.2pt}}

\def\STrow#1{\hbox{#1}\kern-1.35pt}

\def\tri#1#2#3#4#5#6#7#8#9{\matrix{#4\cr
	#3\quad#5\cr #2~\qquad #6\cr #1\quad #9\quad#8\quad#7\cr}}

\def\text#1{\quad\hbox{#1}\quad}

\def\la{\lambda}

\def\rep{\rm representation\ }
\def\reps{\rm representations\ }

\parskip=6pt
\Title{\vbox{\baselineskip12pt
\hbox{LETH-PHY-9/92}\hbox{LAVAL-PHY-23/92}}}
{\vbox {\centerline{Crystallising the depth rule}
\bigskip
\centerline{for WZNW fusion coefficients}}}

\centerline{A.N. Kirillov$^\flat$, P. Mathieu$^\flat$, D.
S\'en\'echal$^\flat$\foot{Present address: D\'epartement de Physique,
Universit\'e de Sherbrooke, Sherbrooke (Qu\'ebec), Canada J1K 2R1} and M.A.
Walton$^\sharp$\foot{Talk given by M.W. at the XIXth International Colloquium
on
Group Theoretical Methods in Physics, June 29 - July 4, 1992, Salamanca,
Spain.}} \vskip.2in \centerline{$^\flat$ \it D\'epartement de Physique,
Universit\'e Laval, Qu\'ebec, Canada G1K 7P4} \smallskip
\centerline{$^\sharp$ \it Physics Department, University of
Lethbridge}\smallskip
\centerline{\it Lethbridge, Alberta, Canada T1K 3M4} \vskip
.3in
\centerline{\bf Abstract}
\bigskip
\noindent
Motivated by a formula (due to Zelobenko) for finite Lie algebra tensor
products, we propose a reformulation of the Gepner-Witten depth rule.
Implementation of this rule remains difficult, however, since the basis states
convenient for calculating tensor product coefficients do not have a
well-defined depth. To avoid this problem, we present a `crystal
depth rule', that gives a lower bound for the minimum level at which
a WZNW fusion appears. The bound seems to be quite accurate for
$su(N>3),$ and for $su(3)$ the rule is proven to be exact.\Date{9/92}

Consider the decomposition of the tensor product of two integrable highest
weight \reps $L(\la)$ and $L(\mu)$ of a finite semi-simple Lie algebra $g$:
\eqn\tp{L(\la)\ \otimes\ L(\mu)\ =\ \oplus_{\nu}\ N_{\la\mu}^{\ \ \ \nu}\
L(\nu)\ \ .}
One has \ref\zref{D.P. Zelobenko, {\it Compact Lie Groups and Their
Representations} (Providence, RI: American Mathematical Society, 1973); Theorem
4, section 78.}:
\eqn\zeq{N_{\la\mu}^{\ \ \ \nu}= {\rm dim}\Big\{|\mu^\prime>\in
L_{\nu-\la}(\mu)  \ \mid\ e_i^{\lambda_i+1}|\mu^\prime>=0\ (i=1,\ldots,{\rm
rank}(g))\Big\}\ ,} where $L_{\nu-\la}(\mu)\subset L(\mu)$ is the subspace of
states of $L(\mu)$ of weight $\nu-\la,$ the $\la_i$ are the Dynkin labels of
the
highest weight $\lambda
,$ and $e_i$ is the raising operator corresponding to the simple root
$\alpha_i$ of $g.$

To each highest weight $\la$ satisfying $k\geq(\la,\theta)$ one can
associate a primary field $\phi_\la$ of the WZNW model of algebra $g$ and
fixed level $k\in{\bf Z}_+.$ The operator products of the primary fields
decompose in a way very similar to \tp. To symbolise this, we write
\eqn\fusion{L(\la)\ \times\ L(\mu)\ =\ \oplus_{\nu}\ N_{\la\mu}^{(k)\ \nu}\
L(\nu)\ \ ,}
calling $\times$ a fusion product, and the $N_{\la\mu}^{(k)\ \nu}$ fusion
coefficients.

The constraints in \zeq\ arise because if one acts on the highest state
$|\la>$ of $L(\la)$ with the lowering operator $f_i$ (corresponding to the
simple
root $\alpha_i$ of $g$) a sufficient number ($\lambda_i+1$) of times, the
result vanishes. These constraints must be obeyed by the fusions \fusion, as
well as by the tensor products \tp. Now the primary field $\phi_\lambda$
creates
the highest state of an integrable highest weight \rep of the untwisted affine
Kac-Moody algebra $\hat g_k$ associated to $g,$ at level $k.$ The affine
algebra
has one extra simple root $\alpha_0$ compared to $g,$ and so for the fusions,
we find an extra condition.

{}From this condition it can be shown \ref\gw{D. Gepner and E. Witten, {\sl
Nucl. Phys.} {\bf B278} (1986) 493.} that a given $L(\nu)$ of the right
hand side of \tp\ will not be part of the right hand side of \fusion\
unless for all $|\mu^\prime>\in L(\mu),\ |\nu^\prime>\in
L(C\nu)$\foot{$L(C\nu)$ is the \rep contragredient to the \rep $L(\nu).$}
obeying  $|\mu^\prime>\otimes|\nu^\prime>\in L(C\la),$ we have
\eqn\dr{f_\theta^{\ \ p} |\mu^\prime>\ \otimes\ f_\theta^{\ \ l} |\nu^\prime>\
=\ 0\ \ ,} for all $p+l\geq k-(\lambda,\theta)+1.$ Here $f_\theta$ is the
lowering operator corresponding to the highest root $\theta$ of $g.$ This
constraint on couplings between 3 WZNW primary fields is known as the depth
rule.

I report here an investigation of the depth rule as a practical method of
calculating the fusion coefficients $N_{\la\mu}^{(k)\ \ \nu}$
\ref\us{A.N. Kirillov, P. Mathieu, D. S\'en\'echal and M.A. Walton, preprint
LAVAL-PHY-20/92(LETH-PHY-2/92), 2/92, to appear in {\sl Nucl. Phys.} {\bf
B}.}. For simplicity, we restricted to $g=su(N).$

First, notice the striking difference between \dr\ and \zeq. The latter allows
the calculation of tensor product coefficients using the simple
Littlewood-Richardson rule. One hopes something similar is possible for fusion
coefficients. Therefore, imitating \zeq,
we used the following `strong' depth rule:
\eqn\sdr{N_{\la\mu}^{(k) \ \nu}= {\rm dim}\Big\{|\mu^\prime>\in
L_{\nu-\la}(\mu)  \ \mid\
e_i^{\lambda_i+1}|\mu^\prime>=f_\theta^{k+1-(\lambda,\theta)}|\mu^\prime>=0
\Big\}\ ,}
where $i=1,\ldots,{\rm rank}(g),$ as in \zeq.

Second, a basis $B_{\nu-\la}(\mu)$ of states $|\mu^\prime>\in L_{\nu-\la}(\mu)$
must be chosen. One might hope to find a basis such that
\eqn\hope{{\rm if}\ \ |\mu^\prime>\in B_{\nu-\lambda}(\mu),\ \ {\rm then\
either}\ \ e_i|\mu^\prime>\in B_{\nu-\la+\alpha_i}(\mu)\ \ {\rm or}\ \
e_i|\mu^\prime>=0.}
That is, a nonvanishing $e_i|\mu^\prime>$ is a single pure element of
$B_{\nu-\la+\alpha_i}(\mu),$ rather than a linear combination of basis states.
This is impossible for all $i,$ however. What we can choose is a basis of
$L_{\nu-\la}(\mu)$ such that by simply dropping some of the states, the
remaining truncated set spans the full set of states obeying the constraints of
\zeq\ (compare with \ref\gz{I.M.
Gelfand  and A.V. Zelevinsky, in the Proceedings of the third seminar, Yurmala,
May 22-24, 1985 (Moscow, Nauka, 1986).}). In fact, there exists a canonical
basis \ref\lu{G. Lusztig, {\sl J. Am. Math. Soc.} {\bf 3} (1990) 447.}
\ref\cb{M. Kashiwara, {\sl Comm. Math. Phys.} {\bf 133} (1990)
249.} parametrised by Gelfand-Tsetlin patterns (GT states) which is a `good'
basis in this sense \ref\om{O. Mathieu, {\sl Geometricae dedicata} {\bf 36}
(1990) 51}, and this property `explains' why the Littlewood-Richardson rule for
$su(N)$ tensor products works.

One needs to decide whether the GT states remain a `good' basis
when the extra constraint of \sdr\ is imposed, i.e. when fusions
are considered instead of tensor products. Unfortunately, we have
not (yet) answered this question.

We did nevertheless make some progress motivated by quantum
groups. It is simple to show using the quantum group
${\cal U}_q(g)$ that the GT
states form a good basis for the calculation of tensor product
coefficients $N_{\la\mu}^{\ \ \ \nu}.$ The tensor product
coefficients of  ${\cal U}_q(g)$ are identical for all generic
$q,$ including $q=1$ (where we have been working) and $q=0.$
Furthermore, the GT patterns label basis states for both values
of $q.$ So we can work at $q=0,$ with Kashiwara's `crystal base'
\cb. These states obey property \hope, as can be seen from
`crystal graphs'. And so it becomes clear that the GT states are `good' for
tensor products.

Now, the GT states at $q=0$ are also `good' with respect to the
full set of constraints of \sdr. But this cannot be used to say
anything about fusion coefficients, because they are identical
to the tensor product coefficients of ${\cal U}_q(g)$ at $q$
equal to a root of unity, i.e. non-generic $q.$ Nevertheless, if
we pretend the states at $q=1$ behave as if they were those at
$q=0,$ we obtain remarkably good results.\foot{There was one complication,
however. The resulting rule was not invariant under
the cyclic permutation of the
three weights $\la, \mu, C\nu$. So, it was necessary to take the strongest of
the three constraints \sdr, with permuted highest weights.} We call the
corresponding depth rule the `crystal depth rule'.

We proved that the crystal depth rule is exact for $su(3),$ and
so provides a simple efficient method for calculating $su(3)$
WZNW fusion coefficients. For $su(N>3),$ we found
evidence that it gives a good lower bound on the minimum level
at which a given coupling appears.

To conclude we present the $su(3)$ crystal depth rule, using
the symmetric form of the Littlewood-Richardson rule for $su(N)$ found by
Berenstein and Zelevinsky \ref\bz{A.D. Berenstein and A.V.
Zelevinsky, `Triple multiplicities for $sl(R+1)$ and spectrum
of the exterior algebra of adjoint representation', Cornell
technical report `90-60 (9/90).}. The tensor product
coefficient $N_{\la\mu}^{\ \ \ C\nu}$ is just the number of
triangles having labels $0\leq a_i\in{\bf Z}$:
\eqn\bzt{\tri{a_1}{a_2~~}{a_3}{a_4}{a_5}{a_6}{a_7}{a_8}{a_9} \text{such that}
\matrix{a_1+a_2=\la_1\cr a_3+a_4=\la_2\cr a_4+a_5=\mu_1\cr
a_6+a_7=\mu_2\cr a_7+a_8=\nu_1\cr a_9+a_1=\nu_2\cr}\qquad
\matrix{a_2+a_3 = a_6+a_8\cr a_3+a_5 = a_9+a_8\cr a_5+a_6
 = a_2+a_9.\cr}}
Suppose the triangle's labels satisfy $a_2\leq{\rm min}(a_5, a_8).$ ( If they
don't we can always rotate the triangle such that they do. ) Then the fusion
coefficient $N_{\la\mu}^{(k)\ C\nu}$ is simply the number of these obeying
$k\geq
k_0=a_4+\nu_1+\nu_2.$

\vskip1cm
\listrefs

\bye